\documentclass[%
 aip,
 jcp,
 amsmath,amssymb,
 reprint,%
]{revtex4-2}

\usepackage{graphicx}
\usepackage{dcolumn}
\usepackage{bm}
\usepackage[utf8]{inputenc}
\usepackage[T1]{fontenc}
\usepackage{mathptmx}
\usepackage{etoolbox}
\usepackage{braket}
\usepackage{url}
\usepackage{color}
\newcounter{num}

\makeatletter
\def\Hline{
  \noalign{\ifnum0=`}\fi\hrule \@height 4.\arrayrulewidth \futurelet\reserved@a\@xhline}

\def\@email#1#2{%
 \endgroup
 \patchcmd{\titleblock@produce}
  {\frontmatter@RRAPformat}
  {\frontmatter@RRAPformat{\produce@RRAP{*#1\href{mailto:#2}{#2}}}\frontmatter@RRAPformat}
  {}{}
}%
\makeatother
\begin{document}

\preprint{}
\title[J. Chem. Phys.]{
Quantification of chirality based on electric toroidal monopole}
\author{A. Inda}
\affiliation{Graduate School of Science, Hokkaido University, Sapporo 060-0810, Japan.}
\author{R. Oiwa}
 \email{}
\affiliation{
  RIKEN Center for Emergent Matter Science (CEMS), Wako 351-0198, Japan.
}
\author{S. Hayami}
\affiliation{Graduate School of Science, Hokkaido University, Sapporo 060-0810, Japan.}
\author{H. M. Yamamoto}
\affiliation{Institute for Molecular Science, Myodaiji, Okazaki 444-8585, Japan.
}
\affiliation{QuaRC, Institute for Molecular Science, Myodaiji, Okazaki 444-8585, Japan.}
\author{H. Kusunose}
\affiliation{QuaRC, Institute for Molecular Science, Myodaiji, Okazaki 444-8585, Japan.}
\affiliation{Department of Physics, Meiji University, Kanagawa 214-8571, Japan.
}

\date{\today}

\begin{abstract}
Chirality ubiquitously appears in nature, however, its quantification remains obscure owing to the lack of microscopic description at the quantum-mechanical level.
We propose a way of evaluating chirality in terms of electric toroidal monopole, a practical entity of time-reversal even pseudoscalar (parity-odd) object reflecting relevant electronic wave functions.
For this purpose, we analyze a twisted methane at the quantum-mechanical level, showing that the electric toroidal monopoles become a quantitative indicator for chirality.
In the twisted methane, we clarify that the handedness of chirality corresponds to the sign of the expectation value of the electric toroidal monopole, and that the most important ingredient is the modulation of the spin-dependent imaginary hopping between the hydrogen atoms, while the relativistic spin-orbit coupling within the carbon atom is irrelevant for chirality.
\end{abstract}

\maketitle

\section{Introduction}
\label{sec:intro}
\begin{figure}[th]
  \includegraphics[width=0.9\hsize]{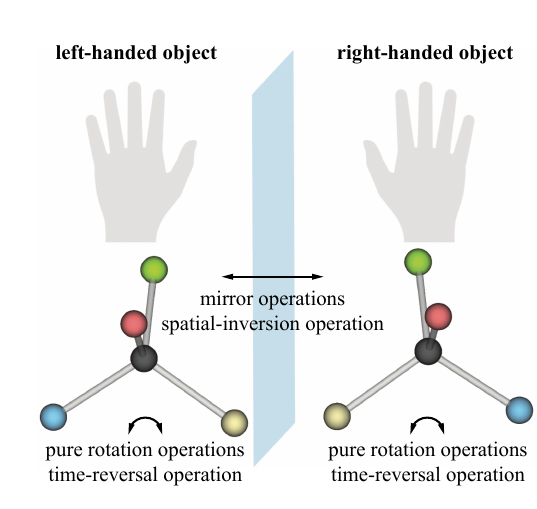}
  \caption{\label{fig:chiral}
  Symmetry aspect of chirality.
  Each enantiomer is invariant from itself by pure rotation and time-reversal operations, while the mirror or spatial-inversion operations transform from one to another.
  }
\end{figure}

Chirality is a profound property of materials, which plays crucial roles in fields of particle physics, condensed matter physics, chemistry, biochemistry, and so on.
A chiral object is characterized by pure rotation operations only, and hence its mirror image is distinguishable to itself~\cite{kelvin1894molecular, kelvin2010baltimore}, such as the right and left hands and asymmetric carbon as shown in Fig.~\ref{fig:chiral}.
In other words, a chiral object does not possess the spatial inversion ($\mathcal{P}$) nor mirror ($\mathcal{M}$) operations.
Moreover, the time-reversal ($\mathcal{T}$) operation is irrelevant to fundamental aspect of chirality.
Barron emphasized these characteristic symmetry properties as ``$\mathcal{T}$-even pseudoscalar'' object~\cite{L.D.Barron_1986_true-chirality,Barron_2004}, and it is consistent with a variety of chirality-induced phenomena, such as Edelstein effect~\cite{T.Yoda_sr_2015_Edelstein,Furukawa2017,T.Furukawa_prr_2021_Edelstein}, electrical magnetochiral effect~\cite{rikken2002magnetochiral,ishizuka2020anomalous}, giant optical activity in nanostructures~\cite{kuwata2005giant}, and non-degenerate chiral phonon excitations~\cite{pine1970direct, ghosh2008origin,PhysRevLett.125.245302,https://doi.org/10.1002/chir.23544, chen2022chiralphonon_diode,K.Ishito_natphys_2023_alpha-HgS, tsunetsugu2023theory,doi:10.7566/JPSJ.92.075002}.
In such research fields, enantioselective synthesis~\cite{christmann2013asymmetric,Barron1986jacs_asymmetric_systhesis}, absolute configuration determination, and characterization of chirality~\cite{bijvoet1951determination, hendry2010ultrasensitive, kelly2020controlling} are particularly important, and they would promote endeavors to unveil the origin of homochirality~\cite{sallembien2022possible, barron2012cosmic}.

Nevertheless, the microscopic description and quantification of chirality have not been fully elucidated yet, i.e., how the degree of chirality affects these phenomena at the quantitative level and which electronic and lattice degrees of freedom play an essential role.
Meanwhile, chirality-induced spin selectivity (CISS)~\cite{B.Gohler_nat_2011_CISS,O.Ben_nat_2017_CISS, K.Michaeli_PNAS_2019_CISS, Suda_natcom_2019_CISS, A.Inui2020prl_CrNb3S6, R.Neeman_ACR_2020_CISS, D.H.Waldeck2021aplmat_CISS, K.Shiota2021prl_disilicide_CISS, A.Kato2022prb_CISS,Nakajima23,PhysRevLett.132.056302} has been investigated extensively, in which electron spins tend to be highly polarized when they are experienced with chiral-materials' environment.
Since the CISS phenomena definitely indicate the importance of electron's spin in chirality, it requires a fundamental understanding of its role at the quantum-mechanical level.

From these circumstances, we aim to elucidate the role of electron's spin in chirality, by evaluating chirality of the twisted methane quantitatively as a prime demonstration.
To this end, we adopt the concept of electric-toroidal monopole (ETM) as a suitable measure of chirality.
The ETM is one of the orthonormal basis in the symmetry-adapted multipole basis (SAMB) set, which has been developed recently for the systematic description of electronic degrees of freedom~\cite{hayami2018microscopic,hayami2018prb_atomic-scale_multipoles,kusunose2020complete,kusunose2022generalization,H.Kusunose_PRB_2023_SAMB}.
The ETM is invariant under any rotation or $\mathcal{T}$ operations, while $\mathcal{P}$ or $\mathcal{M}$ changes the sign of the ETM.
These properties are exactly the same as those of chiral object, i.e., $\mathcal{T}$-even pseudoscalar, and hence it is a suitable quantitative measure for chirality~\cite{hayami2018prb_atomic-scale_multipoles,kusunose2020complete,Oiwa_PhysRevLett.129.116401,J.Kishine_IJC_2022_G0,tsunetsugu2023theory, hayami2023chiral, hayami2023analysis}.

Since the SAMB constitutes a complete basis set in any physical Hilbert space of atoms, molecules, and crystals, it describes any anisotropy of combined electronic degrees of freedom such as charge, orbital, and spin in underlying molecular or crystal structure in a unified manner~\cite{hayami2018microscopic, kusunose2020complete, H.Kusunose_PRB_2023_SAMB}.
It has been utilized not only for systematic identification and classification of electronic order parameters but also for an exploration of cross correlations and nonlinear transport under 32 crystallographic point groups~\cite{hayami2018prb_atomic-scale_multipoles} and 122 magnetic point groups~\cite{yatsushiro2021prb_122}.

In the present study, we demonstrate the quantification of chirality in terms of the ETMs by analyzing the twisted methane CH$_4$, since it has a relatively small number of electronic degrees of freedom in a simple structure.
This example sheds light on the role of the electron's spin in chirality, for instance, we elucidate that the handedness of chirality corresponds to the sign of the expectation values of the ETMs.
In addition, we clarify that in the case of twisted methane, the spin-dependent imaginary hybridizations and the hopping modulations play important roles in activating chirality, whereas the relativistic spin-orbit coupling (SOC) within an atomic site is less important.
Our approach can also be applied to more realistic chiral objects, which brings about new insight of chirality at the quantum-mechanical level, and stimulates microscopic understanding of chirality beyond the symmetry argument.

The remaining part of this paper is organized as follows.
In Sec.~\ref{sec:multipole}, we introduce the SAMB consisting of four-type multipoles and discuss its correspondence to chirality in general.
Then, we investigate the ETMs specifically in the twisted methane as the simplest example in Sec.~\ref{sec:application_to_CH4}.
We show how the ETMs appear when the achiral CH$_{4}$ is twisted to break all the $\mathcal{M}$ symmetries, focusing on the role of bond modulations, SOC, and handedness (twist direction).
Lastly, we summarize the paper in Sec.~\ref{sec:summary}.

\section{Microscopic description of chirality}
\label{sec:multipole}
We introduce the microscopic expressions of chirality based on the SAMB~\cite{hayami2018prb_atomic-scale_multipoles,kusunose2020complete,H.Kusunose_PRB_2023_SAMB}.
In Sec.~\ref{subsec:multipole_basis}, we briefly introduce four types of multipoles constituting the SAMB with different spatial-inversion and time-reversal properties.
Then, we show the expressions of chirality of on-site type and bond type in Sec.~\ref{subsec:G0_expression};
the former represents the spin-dependent imaginary hybridizations between parity distinct orbitals within a single site and the latter represents the spin-dependent imaginary hoppings between different sites.

\subsection{Symmetry-adapted multipole basis}
\label{subsec:multipole_basis}
\begin{table}
  \caption{\label{tab:four_type_multipole}
  Symmetry properties of four types of monopoles and dipoles in terms of the the spatial-inversion $\mathcal{P}$, time-reversal $\mathcal{T}$, and mirror $\mathcal{M}_{\parallel}$ ($\mathcal{M}_{\perp}$) operations parallel (perpendicular) to dipole vector.
  E, M, MT, and ET stand for the electric, magnetic, magnetic toroidal, and electric toroidal multipoles, respectively.}
  \begin{ruledtabular}
    \begin{tabular}{ccccccc}
    Multipole& Notation &$\mathcal{P}$ & $\mathcal{T}$& $\mathcal{M}_{\parallel}$ & $\mathcal{M}_{\perp}$ & Same symmetry property\\
    \Hline
    E monopole &$Q_0$ & $+$ & $+$ & $+$ & $+$ & $\bm{r}\cdot\bm{Q}$\\
    M monopole &$M_0$ & $-$ & $-$ & $-$ & $-$ & $\bm{r}\cdot\bm{M}$\\
    MT monopole &$T_0$ & $+$ & $-$ & $+$ & $+$ & $\bm{r}\cdot\bm{T}$\\
    ET monopole &$G_0$ & $-$ & $+$ & $-$ & $-$ & $\bm{r}\cdot\bm{G}$\\
    \hline
    E dipole &$\bm{Q}$ & $-$ & $+$ & $+$ & $-$ &
      \begin{tabular}{c}
        Position ($\bm{r}$), \\Polarization ($\bm{P}$)
      \end{tabular}\\
    M dipole &$\bm{M}$ & $+$ & $-$ & $-$ & $+$ & Orbital ($\bm{l}$), Spin ($\bm{
    \bm{s}}$)\\
    MT dipole &$\bm{T}$ & $-$ & $-$ & $+$ & $-$ &$\bm{r}\times \bm{M}$\\
    ET dipole &$\bm{G}$ & $+$ & $+$ & $-$ & $+$ & $\bm{r}\times \bm{Q}$, $\bm{l} \times \bm{\sigma}$\\
    \end{tabular}
  \end{ruledtabular}
\end{table}

The SAMB consists of four types of multipoles according to the spatial-inversion and time-reversal parities, $\mathcal{P}$ and $\mathcal{T}$: electric $(\mathcal{P}$, $\mathcal{T})= [(-1)^l,+1]$, magnetic $[(-1)^{l+1},-1]$, magnetic toroidal $[(-1)^l,-1]$, and electric toroidal $[(-1)^{l+1},+1]$ multipoles, where $l$ represents the rank of multipoles.
The electric and magnetic toroidal multipoles describe the polar $\mathcal{P}:(-1)^{l}$ tensor quantity, while the magnetic and electric toroidal multipoles describe the axial $\mathcal{P}:(-1)^{l+1}$ tensor quantity.
As these four types of multipole constitute a complete basis set in any physical Hilbert space~\cite{hayami2018prb_atomic-scale_multipoles,kusunose2020complete,H.Kusunose_PRB_2023_SAMB}, any anisotropic charge, spin, and current distributions are expressed by a linear combination of the SAMBs.
For example, (pseudo)scalar quantities with different $\mathcal{P}$ and $\mathcal{T}$ parities are described by any of four types of monopole basis.
The electric (magnetic) charge with $(\mathcal{P}$, $\mathcal{T})= (+1,+1)$ [$(\mathcal{P}$, $\mathcal{T})= (-1,-1)$] is described by the electric (magnetic) monopole basis $Q_0$ ($M_0$).
In addition, a time-reversal odd ``polar charge'' with $(\mathcal{P}$, $\mathcal{T})= (+1,-1)$ is described by the magnetic toroidal monopole $T_0$~\cite{Hayami_PhysRevB.108.L140409}.
Lastly, a ``chiral charge'' with $(\mathcal{P}$, $\mathcal{T})= (-1,+1)$ is described by the ETM $G_0$~\cite{J.Kishine_IJC_2022_G0}.
Similar arguments also hold for anisotropic components of higher-rank multipoles.
We show the parities of four monopoles $(Q_0, M_0, T_0, G_0)$ and dipoles $(\bm{Q}, \bm{M}, \bm{T}, \bm{G})$ in terms of the $\mathcal{P}$ and $\mathcal{T}$ as well as mirror operations parallel (perpendicular) to dipole vector, $\mathcal{M}_{\parallel}$ ($\mathcal{M}_{\perp}$), in Table~\ref{tab:four_type_multipole}~\cite{Hlinka2014prl_eight-types}.
Note that there is no distinction of $\mathcal{M}_{\parallel}$ and $\mathcal{M}_{\perp}$ for (pseudo)scalar quantity.
In the following subsections, we show the microscopic expressions of the ETMs.

\subsection{On-site and bond chirality}
\label{subsec:G0_expression}

\begin{figure}
  \includegraphics[width=0.3\hsize]{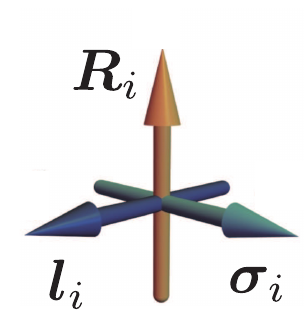}
  \caption{\label{fig:microG0}
  Building block of the ETM.
  It constitutes the site-cluster ETM, where $\bm{R}_{i}\cdot(\bm{l}_{i}\times\bm{\sigma}_{i})$ has
  the same symmetry property of $\mathcal{T}$-even pseudoscalar.
  The yellow arrow represents position vector $\bm{R}_{i}$, and the blue (lightblue) arrow represents the orbital (spin) angular momentum $\bm{l}_{i}$ ($\bm{\sigma}_{i}$) at $\bm{R}_{i}$.
  }
\end{figure}

First, let us discuss the expression of the ETMs within a single atomic site.
The ETM can be expressed as an inner product of polar and axial vectors in general.
The electric dipole $\bm{Q}$ and magnetic toroidal dipole $\bm{T}$ are categorized as the polar vectors, while the magnetic dipole $\bm{M}$ and electric toroidal dipole $\bm{G}$ are categorized as the axial vectors.
By considering the time-reversal property as well, the ETM is represented as $\bm{Q}\cdot\bm{G}$ or $\bm{T}\cdot\bm{M}$.

By considering symmetry properties, the orbital $\bm{l}$ and spin $\bm{s}= {\sigma}/2$ angular momenta correspond to $\bm{M}$, while the imaginary electron transfer between orbitals with different parities (e.g., $s$ and $p$ orbitals) corresponds to $\bm{T}$.
Therefore, one possible expression of the ETM is $\bm{T}\cdot\bm{s}$ or $\bm{T}\cdot\bm{l}$ representing spin or orbital dependent imaginary electron transfer between parity distinct orbitals.
When the transfer occurs within a single atomic site, it is called the atomic ETM, which can be activated in the Hilbert space with parity distinct orbitals, such as $s$-$p$, $p$-$d$, and $d$-$f$ orbitals~\cite{hayami2018prb_atomic-scale_multipoles,kusunose2020complete}.

\begin{figure}
  \includegraphics[width=\hsize]{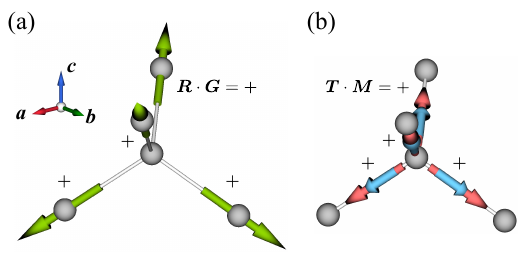}
  \caption{\label{fig:clusterG0}
  Example of ETMs in a tetrahedral cluster, (a) site-cluster type and (b) bond-cluster type.
  The green arrows represent $\mathcal{T}$-even axial vector $\bm{G}$ at the position $\bm{R}$, and the red and blue arrows represent the imaginary spin-dependent hopping ($\mathcal{T}$-odd polar vector)
  $\bm{T}$ and magnetic dipole $\bm{M}$, respectively.}
\end{figure}

Next, we consider the site-cluster ETM.
From the symmetry point of view, the outer product of $\bm{l}$ and $\bm{\sigma}$ corresponds to $\bm{G}$ as an on-site quantity~\cite{Hayami_doi:10.7566/JPSJ.91.113702} as shown in Fig.~\ref{fig:microG0}.
In the case of plural atoms in a molecule, each of which has $\bm{G}_{i}$ at the site $\bm{R}_{i}$, the cluster quantity $\sum_{i}\bm{R}_{i}\cdot\bm{G}_{i}$ represents the ETM as shown in Fig.~\ref{fig:clusterG0}(a).
It is called as the site-cluster ETM, where the spatial distribution of $\bm{R}_{i}\cdot\bm{G}_{i}$ (e.g., $+$ signs) is isotropic.

When the imaginary electron transfer between parity distinct orbitals occurs between the sites $\bm{R}_{i}$ and $\bm{R}_{j}$, the ETM of $(\bm{T}\cdot\bm{\sigma})_{(ij)}$ or $(\bm{T}\cdot\bm{l})_{(ij)}$ is also activated in the $(ij)$ bond as shown in Fig.~\ref{fig:clusterG0}(b), where the spatial distribution of $(\bm{T}\cdot\bm{M})_{(ij)}$ is isotropic.
It corresponds to the spin or orbital dependent imaginary hopping between two sites, and is called as the bond-cluster ETM.

In this way, we can express the ETMs in many different ways depending on the relevant atomic Hilbert space and cluster unit in a molecule.
We adopt the open-source Python library, MultiPie~\cite{H.Kusunose_PRB_2023_SAMB}, to automatically generate the SAMB for specified atomic wave functions and the molecular or crystal structure, focusing on the ETMs as discussed in the following section.

\section{Analysis of chirality in Twisted CH$_4$}
\label{sec:application_to_CH4}

In this section, we analyze the relationship between the chirality and the ETMs by exemplifying the twisted CH$_4$ at the quantum-mechanical level.
Our strategy to unveil chirality is the following: we consider the symmetry lowering of achiral CH$_{4}$ from the point group $T_{\rm d}$ to $D_2$ by twisting the hydrogen atoms as shown in Fig.~\ref{fig:CH4}, and we evaluate quantitatively the activated ETMs in the twisted CH$_4$ molecule.
First, from density-functional theory (DFT) calculation using Quantum ESPRESSO~\cite{Giannozzi_2009_QE}, we introduce a tight-binding model for the CH$_4$ molecule by using the open-source Python library, SymClosestWannier~\cite{SymClosestWannier}, based on MultiPie in Sec.~\ref{subsec:model}.
Next, we discuss possible ETM operators within the above model in Sec.~\ref{subsec:ETM_in_CH4}.
Then, we discuss the effect of the symmetry lowering by considering the proper hopping modulation, and analyze the essential model parameters to activate chirality, i.e. the ETMs.

\subsection{Model}
\label{subsec:model}
\begin{figure}
  \includegraphics[width=\hsize]{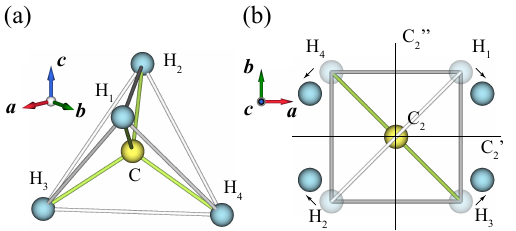}
  \caption{\label{fig:CH4}
  (a) Structure of CH$_4$ under the point group $T_{\rm d}$ where the position of the hydrogen atoms, [H$_{1}$:$(+u,+u,+u)$, H$_{2}$:$(-u,-u,+u)$, H$_{3}$:$(+u,-u,-u)$, H$_{4}$:$(-u,+u,-u)$].
  (b) Twist of H atoms in CH$_{4}$ around $c$ axis with the displacements, [$(+\delta,-\delta,0)$, $(-\delta,+\delta,0)$, $(+\delta,+\delta,0)$, $(-\delta,-\delta,0)$]. It lowers the symmetry from $T_{d}$ to $D_{2}$, where three twofold axes ($C_2, C'_2, C''_2$) remain.}
\end{figure}

We introduce a tight-binding model for the achiral CH$_4$ molecule in a tetrahedral cluster under the point group $T_{\rm d}$, as shown in Fig.~\ref{fig:CH4}(a).
By considering the $s$ and three $p$ orbitals in the C atom and $s$ orbital in the four H atoms, the tight-binding Hamiltonian is given as follows:
\begin{align}
& \mathcal{H}=\mathcal{H}_{\rm pot}+
  \mathcal{H}_{\rm SOC}+
  \mathcal{H}^{\rm Re}_{\rm hop}+
  \mathcal{H}^{\rm Im}_{\rm hop},
  \\&\quad
  \mathcal{H}_{\rm pot}
  =
  \sum_{\sigma=\uparrow,\downarrow}
  V^{s}c^\dagger_{s\sigma}c_{s\sigma}^{}+
  \sum_{\mu=x,y,z} V^p
  c^\dagger_{p_\mu\sigma}c_{p_\mu\sigma}^{}
  +\sum_{i = 1}^{4} V^{\rm H}c^\dagger_{i\sigma}c_{i\sigma}^{},
  \\&\quad
  \mathcal{H}_{\rm SOC}
  = \sum_{\mu\mu'\sigma\sigma'}
  c^{\dagger}_{p_{\mu}\sigma}(\lambda\bm{l}_{\mu\mu'}\cdot \bm{s}_{\sigma\sigma'})c_{p_{\mu'}\sigma'}^{},
  \\&\quad
  \label{eq:Rehop}
  \mathcal{H}^{\rm Re}_{\rm hop}
  =
  c^\dagger (H_t) c^{},
  \\&\quad
  \label{eq:Imhop}
  \mathcal{H}^{\rm Im}_{\rm hop}
  =
  c^\dagger (H^\sigma_t) c,
\end{align}
where $c^\dagger_{s\sigma} (c_{s\sigma}^{})$ and $c^\dagger_{p_\mu\sigma} (c_{p_\mu\sigma}^{})$ for $\mu=x, y, z$ represent the creation (annihilation) operators for the $s$ and $p$ orbitals and spin $\sigma$ in the C atom, respectively, while $c^\dagger_{i\sigma}$ ($c_{i\sigma}$) represents the creation (annihilation) operator for the $s$ orbital and spin $\sigma$ at the H$_i$ ($i=1,2,3,4$) atom.
All the terms in the Hamiltonian belong to the identity representation of $T_{\rm d}$.
$\mathcal{H}_{\rm pot}$ represents the on-site potentials for the $s$ orbital in the C atom [denoted as C($s$)], $V^{s}$, the $p$ orbital in the C atom [denoted as C($p$)], $V^p$, and the $s$ orbital in the H atom, $V^{\rm H}$.
$\mathcal{H}_{\rm SOC}$ represents the relativistic SOC with the magnitude $\lambda$ in the $p$ orbitals of the C atom.
$\mathcal{H}^{\rm Re}_{\rm hop}$ and $\mathcal{H}^{\rm Im}_{\rm hop}$ represent the real and imaginary hoppings between C($s$), C($p$) and H$_{i}$($s$), respectively.
The explicit expressions of the $16\times 16$ matrices $H_{t}$ and $H_{t}^{\sigma}$ are given in Appendix~\ref{app:hop_mat}, where $c^{\dagger}=(c_{s\uparrow}^{\dagger}$, $c_{s\downarrow}^{\dagger}$, $c_{p_{x}\uparrow}^{\dagger}$, $c_{p_{x}\downarrow}^{\dagger}$, $c_{p_{y}\uparrow}^{\dagger}$, $c_{p_{y}\downarrow}^{\dagger}$, $c_{p_{z}\uparrow}^{\dagger}$, $c_{p_{z}\downarrow}^{\dagger}$, $c_{1\uparrow}^{\dagger}$, $c_{1\downarrow}^{\dagger}$, $c_{2\uparrow}^{\dagger}$, $c_{2\downarrow}^{\dagger}$, $c_{3\uparrow}^{\dagger}$, $c_{3\downarrow}^{\dagger}$, $c_{4\uparrow}^{\dagger}$, $c_{4\downarrow}^{\dagger}$).

These hopping terms can be parametrized by five independent parameters $(t^{s\rm H},t^{p\rm H},\tilde{t}^{p\rm H},t^{\rm HH},\tilde{t}^{\rm HH})$; the real hopping between C($s$) and H, the real and imaginary hoppings between C($p$) and H, and between H atoms, respectively.
It is noted that the imaginary hopping should depend on the spin in order to preserve the time-reversal symmetry.

\begin{table}
  \caption{\label{tab:model_parameters}
  The optimized model parameters of CH$_4$ determined by the comparison with DFT calculations with SymClosestWannier~\cite{SymClosestWannier}. }
  \begin{ruledtabular}
  \centering
  \begin{tabular}{cc}
    Parameters & [eV] \\
    \hline
    $V^s$ & -0.867485 \\
    $V^p$ & 0.969553 \\
    $V^{\rm H}$ & -11.2904 \\ \hline
    $\lambda$ & 3.02122 $\times 10^{-4}$\\ \hline
    $t^{s \rm H}$ & -4.30093 \\
    $t^{p \rm H}$ & 0.150131 \\
    $\tilde{t}^{p \rm H}$ & 5.20910 $\times 10^{-3}$\\
    $t^{\rm HH}$ & -8.99125 \\
    $\tilde{t}^{\rm HH}$ & 5.37641 $\times 10^{-3}$ \\
  \end{tabular}
  \end{ruledtabular}
\end{table}

\begin{table}
  \caption{\label{tab:energy_level}
  Energy levels [eV] of CH$_4$ obtained by the DFT calculations with or without SOC.
  The fourth column shows energy levels obtained by the tight-binding (TB) model with the optimized parameters.
  They are classified by the irreducible representation (Irrep.) of $T_{\rm d}$, and the double-group one with the spin degree of freedom in the parenthesis.
  The Fermi energy is set to be zero in all the cases.
  }
  \begin{ruledtabular}
  \centering
  \begin{tabular}{cccc}
  Irrep.& DFT w/o SOC  & DFT w/ SOC & TB Model \\
  \Hline
  $T_2 (\Gamma_7,\Gamma_8)$ &
  \begin{tabular}{c}
    10.4255\\10.4255\\10.4255
  \end{tabular}
  &
  \begin{tabular}{c}
    10.4236 \\ 10.4236 \\ 10.4231
  \end{tabular}
  &
  \begin{tabular}{c}
    10.4236 \\ 10.4236 \\ 10.4231
  \end{tabular}\\
  \hline
  $A_1(\Gamma_6)$ & 8.8735 & 8.8714 & 8.8714\\
  \hline
  $T_2(\Gamma_7,\Gamma_8)$ &
  \begin{tabular}{c}
    0.0\\0.0\\0.0
  \end{tabular}
  &
  \begin{tabular}{c}
    0.0\\0.0\\ -0.0046
  \end{tabular}
  &
  \begin{tabular}{c}
    0.0\\0.0\\ -0.0046
  \end{tabular}\\
  \hline
  $A_1(\Gamma_6)$ & -7.6265 & -7.6279 & -7.6279
  \end{tabular}
  \end{ruledtabular}
\end{table}

As summarized in Table~\ref{tab:model_parameters}, the above model parameters are determined by the comparison with the DFT-based tight-binding model obtained by SymClosestWannier~\cite{SymClosestWannier} Python library, which can determine the model parameters without any iterative calculations based on the Closest Wannier formalism~\cite{ozaki2023closest} and avoid unexpected symmetry lowering due to numerical error with the help of MultiPie~\cite{H.Kusunose_PRB_2023_SAMB}.
The DFT calculations for CH$_{4}$ are performed by using Quantum ESPRESSO~\cite{Giannozzi_2009_QE} with the PBE correlation functional~\cite{J.Perdew1996prl_PAW} and the PAW pseudopotential.
We set the lattice constant to be $a = 10.0~[$\AA$]$ for a simple cubic lattice and $k$ mesh of $1\times 1 \times 1$ in order to deal with an isolated molecule. 
As shown in Fig.~\ref{fig:CH4}(a), the C atom at the origin is surrounded by the four H atoms, [H$_{1}$:$(+u,+u,+u)$, H$_{2}$:$(-u,-u,+u)$, H$_{3}$:$(+u,-u,-u)$, H$_{4}$:$(-u,+u,-u)$], where $u = 0.628~[$\AA$]$. 
We also set the convergence threshold as $10^{-12}~[\rm Ry]$.
We use kinetic energy cutoff for wavefunctions of $160~[\rm Ry]$ and that for charge density and potential of $640~[\rm Ry]$.

As shown in Table~\ref{tab:energy_level}, the comparison of the energy levels between the DFT results with SOC and the tight-binding model gives a perfect agreement.
It is noted that the magnitudes of the SOC $\lambda$ and spin-dependent hoppings $\tilde{t}^{p{\rm H}}$ and $\tilde{t}^{\rm HH}$ are quite small.
This is because both C and H atoms are light elements with considerably small relativistic effect.
It is also consistent with the fact that the difference between DFT energy levels with and without the SOC is small as shown in Table~\ref{tab:energy_level}.

\subsection{ETMs in CH$_4$}
\label{subsec:ETM_in_CH4}
\begin{figure}
  \includegraphics[width=\hsize]{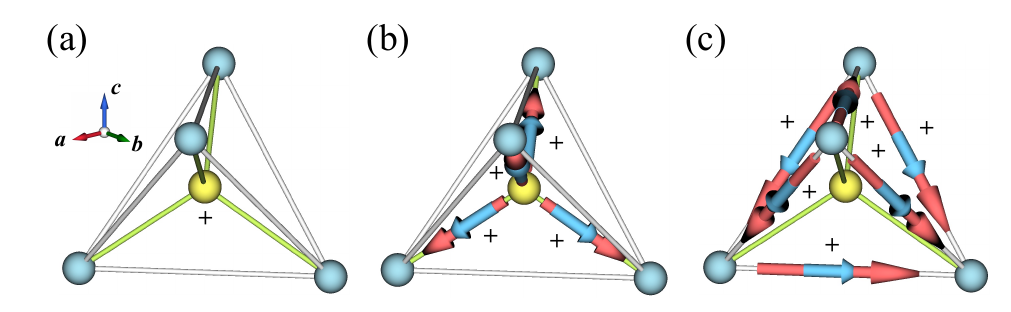}
  \caption{\label{fig:A2}
  ETMs belonging to $A_{2}$ irreducible representation in $T_{\rm d}$.
  (a) the atomic ETM, $G_{0}^{sp}$, (b) the bond-cluster ETMs, $G_{0}^{s\rm H}$ and $G_{0}^{p\rm H}$ on C-H bonds, and (c) the bond-cluster ETM, $G_{0}^{\rm HH}$ on H-H bonds.
}
\end{figure}

In $T_{\rm d}$ symmetry, the ETM and electric-toroidal octupole belong to $A_{2}$ irreducible representation.
In the present Hilbert space, there are four ETMs and one electric-toroidal octupole.
Since H atom has only spin degrees of freedom, i.e., $\bm{l}$ is inactive, and hence all the ETMs are in the form of $\bm{T}\cdot\bm{s}$ as discussed in the previous section.
One is the atomic ETM within the C atom, $G_{0}^{sp}$ as shown in Fig.~\ref{fig:A2}(a).
The remaining three ETMs are the bond-cluster ETMs between C($s$)-H($s$), C($p$)-H($s$), and H($s$)-H($s$), $G_{0}^{s\rm H}$, $G_{0}^{p\rm H}$, and $G_{0}^{\rm HH}$, respectively, as shown in Figs.~\ref{fig:A2}(b) and (c).
The matrix expressions for these ETMs are summarized in Appendix~\ref{app:ETD_mat}.
Note that the expectation values of $G_{0}$ vanish for achiral CH$_{4}$, and they become finite when the molecule is twisted to break all the mirror symmetries as shown later.

\subsection{Effect of symmetry lowering}
\label{subsec:modulation}

Since the ETM belongs to the $A_2$ representation, the symmetry lowering from $T_{\rm d}$ is necessary to activate the ETMs so that it becomes an identity representation.
Although the simplest symmetry lowering is to $T$, as shown in Table~\ref{tab:Td-D2}, such a structural distortion is difficult where the mirror operations must be lost with keeping the three-fold rotation.
Thus, we consider symmetry lowering to $D_2$ by twisting H atoms around the $c$ axis with keeping three two-fold axes, as shown in Fig.~\ref{fig:CH4}(b).

\begin{table}
  \caption{\label{tab:Td-D2}
  Irreducible representations of $T_{\rm d}$ and its subgroups.
  The ETM belongs to the $A_2$ representation.
  }
  \begin{ruledtabular}
  \centering
  \begin{tabular}{cccc}
  $T_{\rm d}$ (CH$_{4}$) & $T$ & $D_2$ (twisted CH$_{4}$) & Remarks\\
  \Hline
  $A_1$ & $A$ & $A$ &\\
  \hline
  $A_2$ & $A$ & $A$ & $G_0$\\
  \hline
  $E$ & $E$ & 2$A$ & \\
  \hline
  $T_1$ & $T$ & $B_1 \oplus B_2 \oplus B_3$ &\\
  \hline
  $T_2$ & $T$ & $B_1 \oplus B_2 \oplus B_3$ &
  \end{tabular}
  \end{ruledtabular}
\end{table}

\begin{figure}
  \includegraphics[width=\hsize]{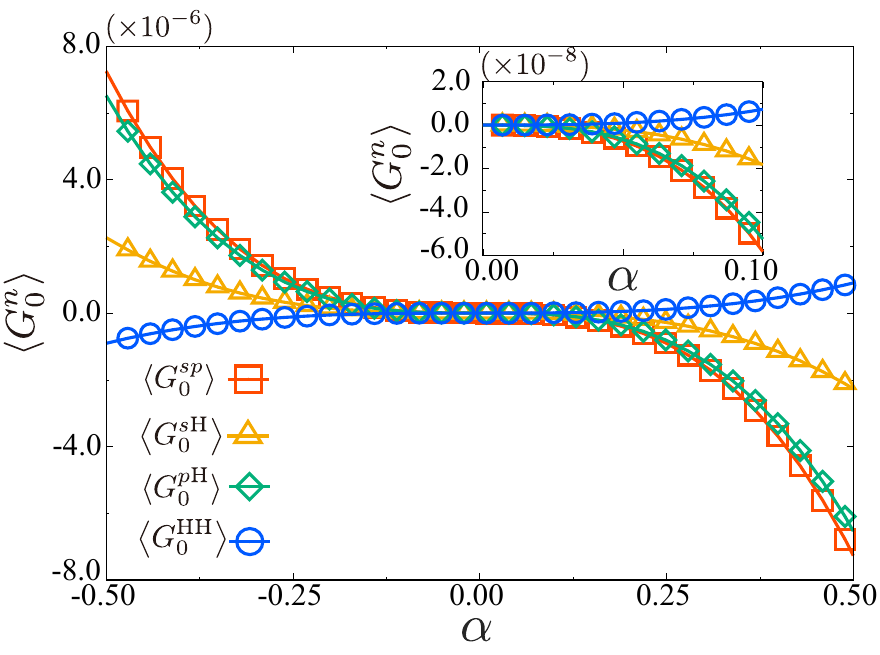}
  \caption{\label{fig:alphafull_dependence} Hopping modulation dependence of $\braket{G^n_0}$ ($n=sp$, $s{\rm H}$, $p{\rm H}$, ${\rm HH}$) for the ground state.
  The other parameters are shown in Table~\ref{tab:model_parameters}.
  The inset represents the enlarged plot around $\alpha=0$.
}
\end{figure}

We take into account the effect of such twisting as the hopping modulations; $t^{p\rm H} \rightarrow t^{p\rm H}\pm\alpha^{p\rm H}$ and $\tilde{t}^{p\rm H}\pm\tilde{\alpha}^{p\rm H}$ for C($p$)-H($s$) bonds, where the upper (lower) sign is for $p_{x}$ ($p_{y}$) orbital.
Moreover, $t^{\rm H H}\rightarrow t^{\rm H H}\pm \alpha^{\rm H H}$ and $\tilde{t}^{\rm H H}\rightarrow \tilde{t}^{\rm H H}\pm \tilde{\alpha}^{\rm H H}$ for H-H bonds, where the upper (lower) sign is for H$_1$-H$_3$ and H$_2$-H$_4$ (H$_1$-H$_4$ and H$_2$-H$_3$) bonds.
Here, $\alpha^{p \rm H},\tilde{\alpha}^{p \rm H},\alpha^{\rm HH},\tilde{\alpha}^{\rm HH}$ represent magnitude of hopping modulations.

Figure~\ref{fig:alphafull_dependence} shows the hopping modulation dependence of the expectation value of the ETMs for the ground state, and we set $\alpha^{p\rm H}=\tilde{\alpha}^{p\rm H}=\alpha^{\rm HH}=\tilde{\alpha}^{\rm HH}=\alpha$ for simplicity.
For $\alpha=0$, $\langle G^n_0 \rangle =0$ ($n=sp$, $s\rm H$, $p\rm H$, ${\rm HH}$) as expected.
By introducing $\alpha$, all $\langle G^n_0 \rangle $ become finite, whose magnitudes become larger as $|\alpha|$ increases.
Therefore, $\braket{G_{0}^{n}}$ plays a quantitative measure for chirality beyond the symmetry argument.
Furthermore, the sign of $\braket{G^n_0}$ is reversed when the twisted direction is reversed as $\alpha \to -\alpha$.
This indicates that the sign of $\braket{G^n_0}$ represents the handedness of chirality.

\begin{figure}
  \includegraphics[width=\hsize]{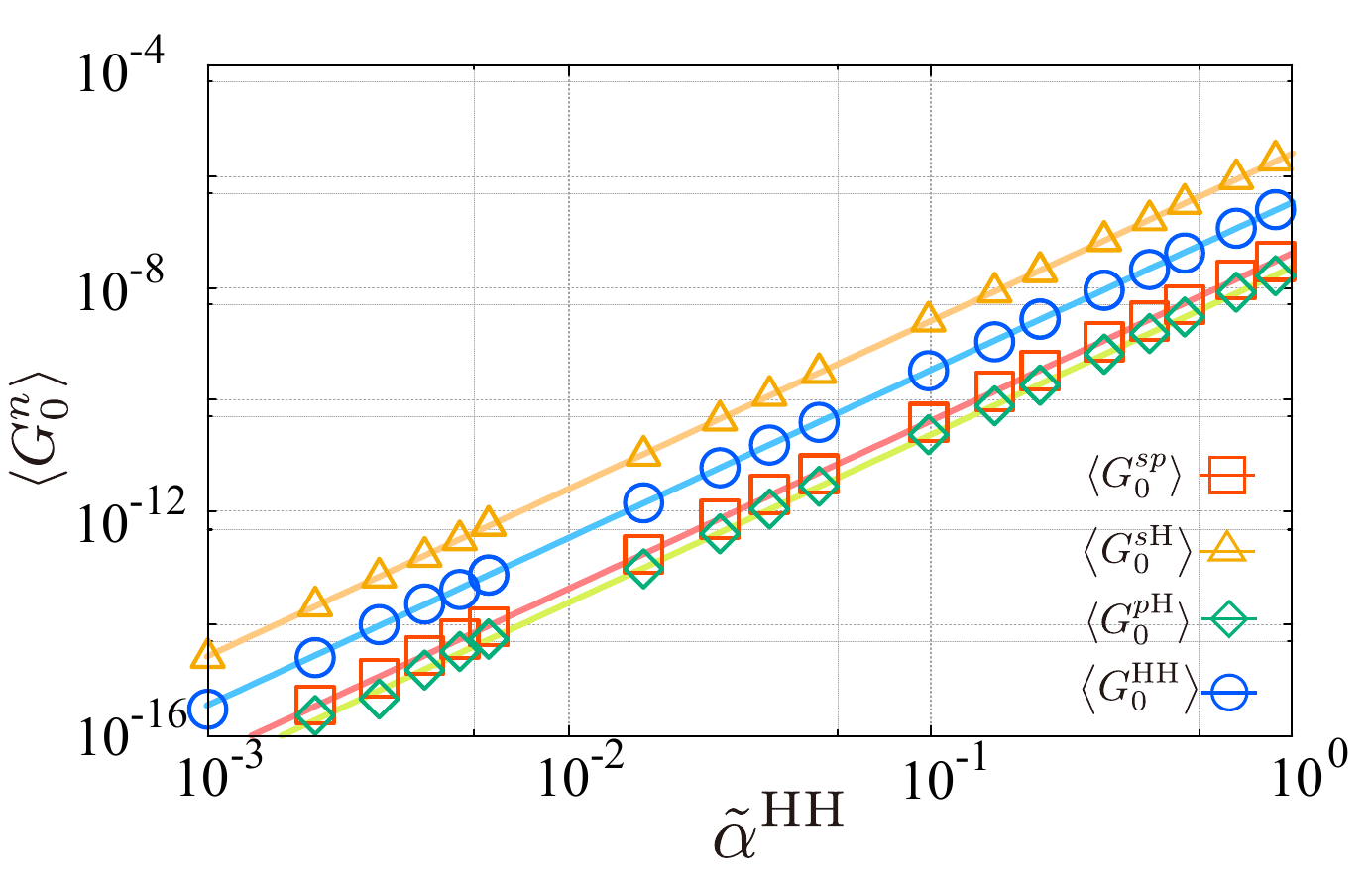}
  \caption{\label{fig:alpha9_order}
  $\tilde{\alpha}^{\rm HH}$ dependence of $\braket{G^n_0}$ ($n=sp$, $s\rm H$, $p\rm H$, ${\rm HH}$) for the ground state.
  The solid lines represent $c_{n} (\tilde{\alpha}^{\rm HH})^3$ where $c_{sp} = 4.25275\times 10^{-8}$, $c_{s\rm H} = 2.59345\times 10^{-6}$, $c_{p\rm H} = 2.40177\times 10^{-8}$, and $c_{\rm HH} =3.44564\times 10^{-7}$.
  The other parameters are chosen as $\alpha^{p \rm H}=\tilde{\alpha}^{p \rm H}=\alpha^{\rm HH}=0$.
}
\end{figure}
\begin{figure}
  \includegraphics[width=\hsize]{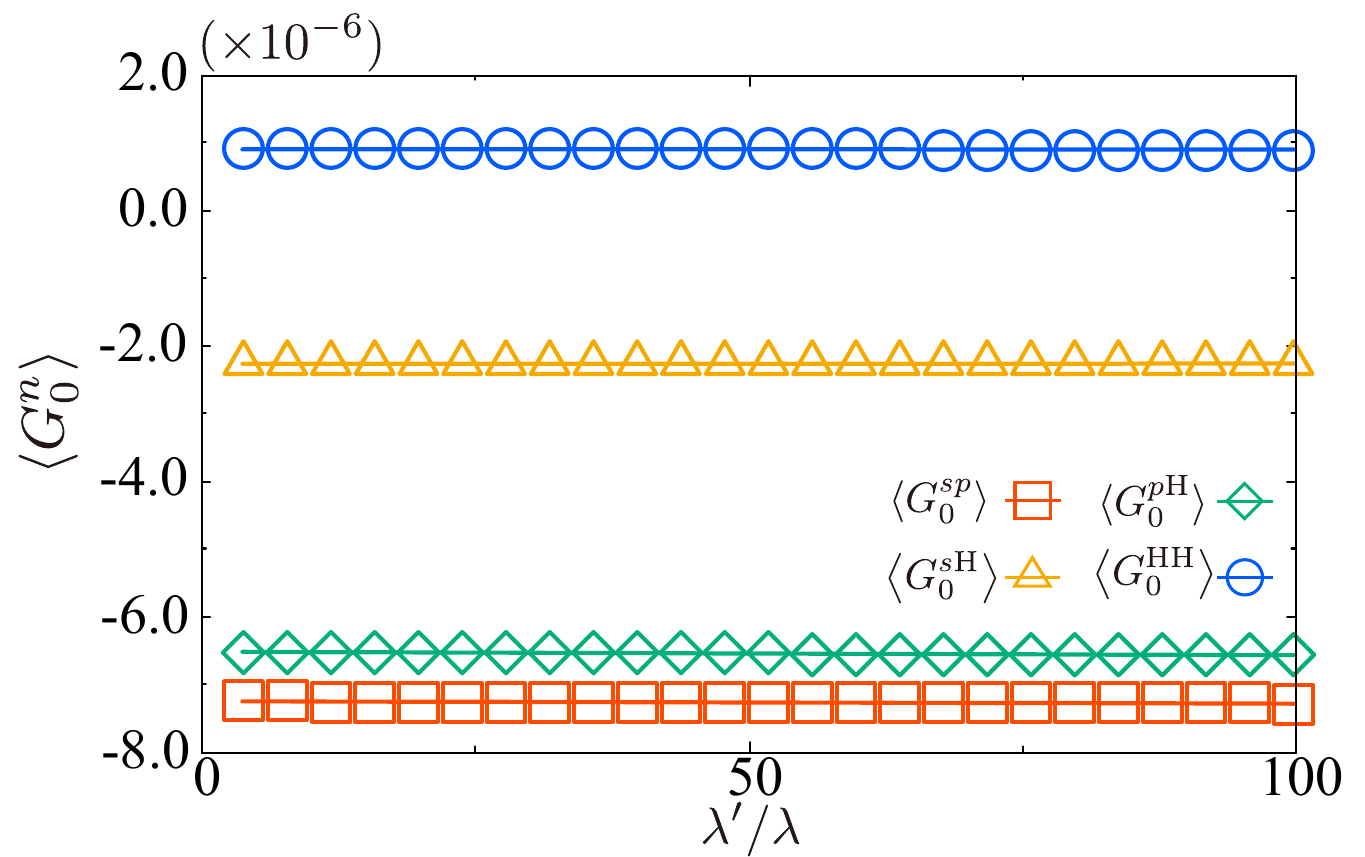}
  \caption{\label{fig:SOC_dependence}
  $\lambda$ dependence of $\braket{G^n_0}$ ($n=sp$, $s\rm H$, $p\rm H$, ${\rm HH}$) for the ground state.
  The other parameters are chosen as $\alpha^{p \rm H}=\tilde{\alpha}^{p \rm H}=\alpha^{\rm HH}=\tilde{\alpha}^{\rm HH}=0.5$.
}
\end{figure}

\begin{table}
  \caption{\label{tab:essential_parameters}
  Essential model parameters (denoted by $\checkmark$) to activate finite $\langle G^n_0 \rangle$ ($n=sp,s{\rm H},p{\rm H},{\rm HH}$).
  }
  \begin{ruledtabular}
  \begin{tabular}{ccccccccccc}
  $G_0$ & $\lambda$ & $\alpha^{p\rm H}$ & $\tilde{\alpha}^{p\rm H}$ & $\alpha^{\rm HH}$ & $\tilde{\alpha}^{{\rm HH}}$& $t^{s\rm H}$ & $t^{p\rm H}$ & $\tilde{t}^{p{\rm H}}$& $t^{\rm HH}$ & $\tilde{t}^{\rm HH}$ \\
  \hline
  $G^{sp}_0$     & -- & -- & -- & -- & \checkmark & -- & -- & \checkmark & -- & -- \\
  $G^{s\rm H}_0$ & -- & -- & -- & -- & \checkmark & -- & -- &  -- & -- & -- \\
  $G^{p\rm H}_0$ & -- & -- & -- & -- & \checkmark & -- & -- & \checkmark & -- & --\\
  $G^{\rm HH}_0$ & -- & -- & -- & -- & \checkmark & -- & -- & -- & -- & -- \\
  \end{tabular}
  \end{ruledtabular}
\end{table}

To extract the important model parameters to activate $\braket{G_{0}^{n}}$, we perform the expansion method following Refs.~\onlinecite{hayami2020prb_bottom-up, R.Oiwa2022jpsj_nonl-respo}.
The results are shown in Table~\ref{tab:essential_parameters}, where $\tilde{\alpha}^{\rm HH}$ is essential to activate all the ETMs; the lowest-order contributions in the expansion of $\braket{G_{0}^{n}}$ are proportional to $(\tilde{\alpha}^{\rm HH})^3$.
It is noted that there is a slight difference of the essential model parameters for ($G^{sp}_0$, $G^{p\rm H}_0$) and ($G^{s\rm H}_0$, $G^{\rm HH}_0$); the former needs $\tilde{t}^{p\rm H}$, while the latter does not.
In addition, $\lambda$ is not necessary for all $\langle G^n_0 \rangle$.

We confirm that all $\langle G^n_0 \rangle $ are proportional to $(\tilde{\alpha}^{\rm HH})^3$ as shown in Fig.~\ref{fig:alpha9_order}, where we set $\alpha^{\rm pH}=\tilde{\alpha}^{\rm pH}=\alpha^{\rm HH}=0$ for simplicity. 
$\langle G^n_0 \rangle $ are proportional to the cubic of $\tilde{\alpha}^{\rm HH}$ because a three-dimentional structural chirality can only be expressed by at least three different hoppings among $s$ orbital of H atoms.
However, chirality can also be expressed by one or two hoppings if the base atomic orbitals have orbital degrees of freedom, which will be a topic of future study. 
In addition, we investigate the effect of $\lambda$ by replacing $\lambda\to\lambda'$.
Figure~\ref{fig:SOC_dependence} shows $\lambda'$ dependence of $\langle G^n_0 \rangle $.
$\langle G^n_0 \rangle $ are almost independent of $\lambda'$.
These results clearly indicate that the modulation of the spin-dependent imaginary hopping between H atoms, $\tilde{\alpha}^{\rm HH}$, is the most important parameter to acquire the chiral nature in twisted CH$_{4}$.

\section{Summary}
\label{sec:summary}
In summary, we have demonstrated by using the twisted CH$_{4}$ that the ETMs are useful quantitative indicators for chirality.
Based on the tight-binding model, in which the model parameters are determined from the DFT calculations with the help of SymClosestWannier and MultiPie Python libraries, we have shown that there are one atomic and three bond-cluster ETMs in CH$_{4}$.
They become finite when the symmetry of CH$_{4}$ lowers from $T_{\rm d}$ to $D_{2}$, which are activated microscopically through the modulation of the spin-dependent imaginary hopping amplitudes.
We have also shown that the sign of the expectation values of the ETMs represents the handedness of chirality as expected.
In addition, we have elucidated that the spin-dependent imaginary hopping between H atoms is the most important parameter, while the relativistic SOC within an atom is irrelevant for chirality in the case of CH$_{4}$.

The present approach can be straightforwardly applied to other chiral molecules and crystals such as CrNb$_3$S$_6$~\cite{T.Morita_ssc_1982_CrNb3S6,T.Miyadai_1983_jpsj_CrNb3S6,Y.Togawa_jpsj_2016_CrNb3S6,A.Inui2020prl_CrNb3S6}, $\alpha$-HgS~\cite{K.Ishito_natphys_2023_alpha-HgS}, NbSi$_2$~\cite{K.Shiota2021prl_disilicide_CISS, H.Shishido_apl_2021_NbSi2-TaSi2}, TaSi$_2$~\cite{K.Shiota2021prl_disilicide_CISS, H.Shishido_apl_2021_NbSi2-TaSi2}, and so on.
The quantitative comparison of chirality in terms of ETMs would provide a deep understanding of chirality at the quantum-mechanical level.

\begin{acknowledgments}
  This research was supported by JSPS KAKENHI Grants Numbers JP21H01037, JP22H04468, JP22H00101, JP22H01183, JP23H04869, JP23K03288, JP23H00091, and by JST PRESTO (JPMJPR20L8) and JST CREST (JPMJCR23O4), and the grants of Special Project (IMS program 23IMS1101), and OML Project (NINS program No. OML012301) by the National Institutes of Natural Sciences.
\end{acknowledgments}
\appendix

\begin{widetext}
\section{Matrix representation of hopping Hamiltonian}
\label{app:hop_mat}
The explicit expressions of the hopping Hamiltonian in Eqs.~(\ref{eq:Rehop}) and (\ref{eq:Imhop}) are given as follows. The blocks of C($s$), C($p$), and H($s$) are separated by solid lines.

\begin{align}
  H_t
  &=
  t^{s\rm H} \mathbb{Z}^{s\rm H}_t + t^{p\rm H} \mathbb{Z}^{p\rm H}_t + t^{\rm HH} \mathbb{Z}^{\rm HH}_t, \\
  H^\sigma_t
  &=
  \tilde{t}^{p\rm H} \mathbb{Z}^{p\rm H}_\sigma + \tilde{t}^{\rm HH} \mathbb{Z}^{\rm HH}_\sigma,
  \end{align}
  \begin{align}
    \mathbb{Z}^{s\rm H}_t
    &=
    \frac{1}{4}
    \left[
      \begin{array}{cc|cccccc|cccccccc}
        0 & 0 & 0 & 0 & 0 & 0 & 0 & 0 & 1 & 0 & 1 & 0 & 1 & 0 & 1 & 0 \\
        0 & 0 & 0 & 0 & 0 & 0 & 0 & 0 & 0 & 1 & 0 & 1 & 0 & 1 & 0 & 1 \\\hline
        0 & 0 & 0 & 0 & 0 & 0 & 0 & 0 & 0 & 0 & 0 & 0 & 0 & 0 & 0 & 0 \\
        0 & 0 & 0 & 0 & 0 & 0 & 0 & 0 & 0 & 0 & 0 & 0 & 0 & 0 & 0 & 0 \\
        0 & 0 & 0 & 0 & 0 & 0 & 0 & 0 & 0 & 0 & 0 & 0 & 0 & 0 & 0 & 0 \\
        0 & 0 & 0 & 0 & 0 & 0 & 0 & 0 & 0 & 0 & 0 & 0 & 0 & 0 & 0 & 0 \\
        0 & 0 & 0 & 0 & 0 & 0 & 0 & 0 & 0 & 0 & 0 & 0 & 0 & 0 & 0 & 0 \\
        0 & 0 & 0 & 0 & 0 & 0 & 0 & 0 & 0 & 0 & 0 & 0 & 0 & 0 & 0 & 0 \\\hline
        1 & 0 & 0 & 0 & 0 & 0 & 0 & 0 & 0 & 0 & 0 & 0 & 0 & 0 & 0 & 0 \\
        0 & 1 & 0 & 0 & 0 & 0 & 0 & 0 & 0 & 0 & 0 & 0 & 0 & 0 & 0 & 0 \\
        1 & 0 & 0 & 0 & 0 & 0 & 0 & 0 & 0 & 0 & 0 & 0 & 0 & 0 & 0 & 0 \\
        0 & 1 & 0 & 0 & 0 & 0 & 0 & 0 & 0 & 0 & 0 & 0 & 0 & 0 & 0 & 0 \\
        1 & 0 & 0 & 0 & 0 & 0 & 0 & 0 & 0 & 0 & 0 & 0 & 0 & 0 & 0 & 0 \\
        0 & 1 & 0 & 0 & 0 & 0 & 0 & 0 & 0 & 0 & 0 & 0 & 0 & 0 & 0 & 0 \\
        1 & 0 & 0 & 0 & 0 & 0 & 0 & 0 & 0 & 0 & 0 & 0 & 0 & 0 & 0 & 0 \\
        0 & 1 & 0 & 0 & 0 & 0 & 0 & 0 & 0 & 0 & 0 & 0 & 0 & 0 & 0 & 0 \\
      \end{array}
  \right],
\end{align}
\begin{align}
  \mathbb{Z}^{p\rm H}_t&=\frac{1}{4\sqrt{3}}
  \left[
    \begin{array}{cc|cccccc|cccccccc}
      0 & 0 & 0 & 0 & 0 & 0 & 0 & 0 & 0 & 0 & 0 & 0 & 0 & 0 & 0 & 0 \\
      0 & 0 & 0 & 0 & 0 & 0 & 0 & 0 & 0 & 0 & 0 & 0 & 0 & 0 & 0 & 0 \\\hline
      0 & 0 & 0 & 0 & 0 & 0 & 0 & 0 & 1 & 0 & -1 & 0 & 1 & 0 & -1 & 0 \\
      0 & 0 & 0 & 0 & 0 & 0 & 0 & 0 & 0 & 1 & 0 & -1 & 0 & 1 & 0 & -1 \\
      0 & 0 & 0 & 0 & 0 & 0 & 0 & 0 & 1 & 0 & -1 & 0 & -1 & 0 & 1 & 0 \\
      0 & 0 & 0 & 0 & 0 & 0 & 0 & 0 & 0 & 1 & 0 & -1 & 0 & -1 & 0 & 1 \\
      0 & 0 & 0 & 0 & 0 & 0 & 0 & 0 & 1 & 0 & 1 & 0 & -1 & 0 & -1 & 0 \\
      0 & 0 & 0 & 0 & 0 & 0 & 0 & 0 & 0 & 1 & 0 & 1 & 0 & -1 & 0 & -1 \\\hline
      0 & 0 & 1 & 0 & 1 & 0 & 1 & 0 & 0 & 0 & 0 & 0 & 0 & 0 & 0 & 0 \\
      0 & 0 & 0 & 1 & 0 & 1 & 0 & 1 & 0 & 0 & 0 & 0 & 0 & 0 & 0 & 0 \\
      0 & 0 & -1 & 0 & -1 & 0 & 1 & 0 & 0 & 0 & 0 & 0 & 0 & 0 & 0 & 0 \\
      0 & 0 & 0 & -1 & 0 & -1 & 0 & 1 & 0 & 0 & 0 & 0 & 0 & 0 & 0 & 0 \\
      0 & 0 & 1 & 0 & -1 & 0 & -1 & 0 & 0 & 0 & 0 & 0 & 0 & 0 & 0 & 0 \\
      0 & 0 & 0 & 1 & 0 & -1 & 0 & -1 & 0 & 0 & 0 & 0 & 0 & 0 & 0 & 0 \\
      0 & 0 & -1 & 0 & 1 & 0 & -1 & 0 & 0 & 0 & 0 & 0 & 0 & 0 & 0 & 0 \\
      0 & 0 & 0 & -1 & 0 & 1 & 0 & -1 & 0 & 0 & 0 & 0 & 0 & 0 & 0 & 0 \\
    \end{array}
  \right],
\end{align}
\begin{align}
  \mathbb{Z}^{p\rm H}_\sigma&=\frac{1}{4\sqrt{6}}
  \left[
    \begin{array}{cc|cccccc|cccccccc}
      0 & 0 & 0 & 0 & 0 & 0 & 0 & 0 & 0 & 0 & 0 & 0 & 0 & 0 & 0 & 0 \\
      0 & 0 & 0 & 0 & 0 & 0 & 0 & 0 & 0 & 0 & 0 & 0 & 0 & 0 & 0 & 0 \\\hline
      0 & 0 & 0 & 0 & 0 & 0 & 0 & 0 & -i & 1 & i & 1 & i & -1 & -i & -1 \\
      0 & 0 & 0 & 0 & 0 & 0 & 0 & 0 & -1 & i & -1 & -i & 1 & -i & 1 & i \\
      0 & 0 & 0 & 0 & 0 & 0 & 0 & 0 & i & -i & -i & -i & i & i & -i & i \\
      0 & 0 & 0 & 0 & 0 & 0 & 0 & 0 & -i & -i & -i & i & i & -i & i & i \\
      0 & 0 & 0 & 0 & 0 & 0 & 0 & 0 & 0 & -1+i & 0 & 1-i & 0 & -1-i & 0 & 1+i \\
      0 & 0 & 0 & 0 & 0 & 0 & 0 & 0 & 1+i & 0 & -1-i & 0 & 1-i & 0 & -1+i & 0 \\\hline
      0 & 0 & i & -1 & -i & i & 0 & 1-i & 0 & 0 & 0 & 0 & 0 & 0 & 0 & 0 \\
      0 & 0 & 1 & -i & i & i & -1-i & 0 & 0 & 0 & 0 & 0 & 0 & 0 & 0 & 0 \\
      0 & 0 & -i & -1 & i & i & 0 & -1+i & 0 & 0 & 0 & 0 & 0 & 0 & 0 & 0 \\
      0 & 0 & 1 & i & i & -i & 1+i & 0 & 0 & 0 & 0 & 0 & 0 & 0 & 0 & 0 \\
      0 & 0 & -i & 1 & -i & -i & 0 & 1+i & 0 & 0 & 0 & 0 & 0 & 0 & 0 & 0 \\
      0 & 0 & -1 & i & -i & i & -1+i & 0 & 0 & 0 & 0 & 0 & 0 & 0 & 0 & 0 \\
      0 & 0 & i & 1 & i & -i & 0 & -1-i & 0 & 0 & 0 & 0 & 0 & 0 & 0 & 0 \\
      0 & 0 & -1 & -i & -i & -i & 1-i & 0 & 0 & 0 & 0 & 0 & 0 & 0 & 0 & 0 \\
    \end{array}
  \right],
\end{align}
\begin{align}
  \mathbb{Z}^{\rm HH}_t&= \frac{1}{2\sqrt{6}}
  \left[
    \begin{array}{cc|cccccc|cccccccc}
      0 & 0 & 0 & 0 & 0 & 0 & 0 & 0 & 0 & 0 & 0 & 0 & 0 & 0 & 0 & 0 \\
      0 & 0 & 0 & 0 & 0 & 0 & 0 & 0 & 0 & 0 & 0 & 0 & 0 & 0 & 0 & 0 \\ \hline
      0 & 0 & 0 & 0 & 0 & 0 & 0 & 0 & 0 & 0 & 0 & 0 & 0 & 0 & 0 & 0 \\
      0 & 0 & 0 & 0 & 0 & 0 & 0 & 0 & 0 & 0 & 0 & 0 & 0 & 0 & 0 & 0 \\
      0 & 0 & 0 & 0 & 0 & 0 & 0 & 0 & 0 & 0 & 0 & 0 & 0 & 0 & 0 & 0 \\
      0 & 0 & 0 & 0 & 0 & 0 & 0 & 0 & 0 & 0 & 0 & 0 & 0 & 0 & 0 & 0 \\
      0 & 0 & 0 & 0 & 0 & 0 & 0 & 0 & 0 & 0 & 0 & 0 & 0 & 0 & 0 & 0 \\
      0 & 0 & 0 & 0 & 0 & 0 & 0 & 0 & 0 & 0 & 0 & 0 & 0 & 0 & 0 & 0 \\ \hline
      0 & 0 & 0 & 0 & 0 & 0 & 0 & 0 & 0 & 0 & 1 & 0 & 1 & 0 & 1 & 0 \\
      0 & 0 & 0 & 0 & 0 & 0 & 0 & 0 & 0 & 0 & 0 & 1 & 0 & 1 & 0 & 1 \\
      0 & 0 & 0 & 0 & 0 & 0 & 0 & 0 & 1 & 0 & 0 & 0 & 1 & 0 & 1 & 0 \\
      0 & 0 & 0 & 0 & 0 & 0 & 0 & 0 & 0 & 1 & 0 & 0 & 0 & 1 & 0 & 1 \\
      0 & 0 & 0 & 0 & 0 & 0 & 0 & 0 & 1 & 0 & 1 & 0 & 0 & 0 & 1 & 0 \\
      0 & 0 & 0 & 0 & 0 & 0 & 0 & 0 & 0 & 1 & 0 & 1 & 0 & 0 & 0 & 1 \\
      0 & 0 & 0 & 0 & 0 & 0 & 0 & 0 & 1 & 0 & 1 & 0 & 1 & 0 & 0 & 0 \\
      0 & 0 & 0 & 0 & 0 & 0 & 0 & 0 & 0 & 1 & 0 & 1 & 0 & 1 & 0 & 0 \\
    \end{array}
  \right],\\
  \mathbb{Z}^{\rm HH}_\sigma&= \frac{1}{4\sqrt{3}}
  \left[
    \begin{array}{cc|cccccc|cccccccc}
      0 & 0 & 0 & 0 & 0 & 0 & 0 & 0 & 0 & 0 & 0 & 0 & 0 & 0 & 0 & 0 \\
      0 & 0 & 0 & 0 & 0 & 0 & 0 & 0 & 0 & 0 & 0 & 0 & 0 & 0 & 0 & 0 \\ \hline
      0 & 0 & 0 & 0 & 0 & 0 & 0 & 0 & 0 & 0 & 0 & 0 & 0 & 0 & 0 & 0 \\
      0 & 0 & 0 & 0 & 0 & 0 & 0 & 0 & 0 & 0 & 0 & 0 & 0 & 0 & 0 & 0 \\
      0 & 0 & 0 & 0 & 0 & 0 & 0 & 0 & 0 & 0 & 0 & 0 & 0 & 0 & 0 & 0 \\
      0 & 0 & 0 & 0 & 0 & 0 & 0 & 0 & 0 & 0 & 0 & 0 & 0 & 0 & 0 & 0 \\
      0 & 0 & 0 & 0 & 0 & 0 & 0 & 0 & 0 & 0 & 0 & 0 & 0 & 0 & 0 & 0 \\
      0 & 0 & 0 & 0 & 0 & 0 & 0 & 0 & 0 & 0 & 0 & 0 & 0 & 0 & 0 & 0 \\ \hline
      0 & 0 & 0 & 0 & 0 & 0 & 0 & 0 & 0 & 0 & 0 & 1-i & i & -1 & -i & i \\
      0 & 0 & 0 & 0 & 0 & 0 & 0 & 0 & 0 & 0 & -1-i & 0 & 1 & -i & i & i \\
      0 & 0 & 0 & 0 & 0 & 0 & 0 & 0 & 0 & -1+i & 0 & 0 & -i & -i & i & 1 \\
      0 & 0 & 0 & 0 & 0 & 0 & 0 & 0 & 1+i & 0 & 0 & 0 & -i & i & -1 & -i \\
      0 & 0 & 0 & 0 & 0 & 0 & 0 & 0 & -i & 1 & i & i & 0 & 0 & 0 & -1-i \\
      0 & 0 & 0 & 0 & 0 & 0 & 0 & 0 & -1 & i & i & -i & 0 & 0 & 1-i & 0 \\
      0 & 0 & 0 & 0 & 0 & 0 & 0 & 0 & i & -i & -i & -1 & 0 & 1+i & 0 & 0 \\
      0 & 0 & 0 & 0 & 0 & 0 & 0 & 0 & -i & -i & 1 & i & -1+i & 0 & 0 & 0 \\
    \end{array}
  \right].
\end{align}

\section{Matrix representation of ETMs}
\label{app:ETD_mat}
The explicit expressions of the ETMs in CH$_4$ as follows.
\begin{align}
  G^{sp}_0
  &=c^{\dagger} \frac{1}{\sqrt{6}}
  \left[
    \begin{array}{cc|cccccc|cccccccc}
      0 & 0 & 0 & i & 0 & 1 & i & 0 & 0 & 0 & 0 & 0 & 0 & 0 & 0 & 0 \\
      0 & 0 & i & 0 & -1 & 0 & 0 & -i & 0 & 0 & 0 & 0 & 0 & 0 & 0 & 0 \\ \hline
      0 & -i & 0 & 0 & 0 & 0 & 0 & 0 & 0 & 0 & 0 & 0 & 0 & 0 & 0 & 0 \\
      -i & 0 & 0 & 0 & 0 & 0 & 0 & 0 & 0 & 0 & 0 & 0 & 0 & 0 & 0 & 0 \\
      0 & -1 & 0 & 0 & 0 & 0 & 0 & 0 & 0 & 0 & 0 & 0 & 0 & 0 & 0 & 0 \\
      1 & 0 & 0 & 0 & 0 & 0 & 0 & 0 & 0 & 0 & 0 & 0 & 0 & 0 & 0 & 0 \\
      -i & 0 & 0 & 0 & 0 & 0 & 0 & 0 & 0 & 0 & 0 & 0 & 0 & 0 & 0 & 0 \\
      0 & i & 0 & 0 & 0 & 0 & 0 & 0 & 0 & 0 & 0 & 0 & 0 & 0 & 0 & 0 \\
      0 & 0 & 0 & 0 & 0 & 0 & 0 & 0 & 0 & 0 & 0 & 0 & 0 & 0 & 0 & 0 \\ \hline
      0 & 0 & 0 & 0 & 0 & 0 & 0 & 0 & 0 & 0 & 0 & 0 & 0 & 0 & 0 & 0 \\
      0 & 0 & 0 & 0 & 0 & 0 & 0 & 0 & 0 & 0 & 0 & 0 & 0 & 0 & 0 & 0 \\
      0 & 0 & 0 & 0 & 0 & 0 & 0 & 0 & 0 & 0 & 0 & 0 & 0 & 0 & 0 & 0 \\
      0 & 0 & 0 & 0 & 0 & 0 & 0 & 0 & 0 & 0 & 0 & 0 & 0 & 0 & 0 & 0 \\
      0 & 0 & 0 & 0 & 0 & 0 & 0 & 0 & 0 & 0 & 0 & 0 & 0 & 0 & 0 & 0 \\
      0 & 0 & 0 & 0 & 0 & 0 & 0 & 0 & 0 & 0 & 0 & 0 & 0 & 0 & 0 & 0 \\
      0 & 0 & 0 & 0 & 0 & 0 & 0 & 0 & 0 & 0 & 0 & 0 & 0 & 0 & 0 & 0 \\
      \end{array}
      \right]c^{},
\end{align}
\begin{align}
  G^{s\rm H}_0 &= c^{\dagger}\frac{1}{4\sqrt{3}}
  \left[
    \begin{array}{cc|cccccc|cccccccc}
      0 & 0 & 0 & 0 & 0 & 0 & 0 & 0 & -i & -1-i & -i & 1+i & i & 1-i & i & -1+i \\
      0 & 0 & 0 & 0 & 0 & 0 & 0 & 0 & 1-i & i & -1+i & i & -1-i & -i & 1+i & -i \\ \hline
      0 & 0 & 0 & 0 & 0 & 0 & 0 & 0 & 0 & 0 & 0 & 0 & 0 & 0 & 0 & 0 \\
      0 & 0 & 0 & 0 & 0 & 0 & 0 & 0 & 0 & 0 & 0 & 0 & 0 & 0 & 0 & 0 \\
      0 & 0 & 0 & 0 & 0 & 0 & 0 & 0 & 0 & 0 & 0 & 0 & 0 & 0 & 0 & 0 \\
      0 & 0 & 0 & 0 & 0 & 0 & 0 & 0 & 0 & 0 & 0 & 0 & 0 & 0 & 0 & 0 \\
      0 & 0 & 0 & 0 & 0 & 0 & 0 & 0 & 0 & 0 & 0 & 0 & 0 & 0 & 0 & 0 \\
      0 & 0 & 0 & 0 & 0 & 0 & 0 & 0 & 0 & 0 & 0 & 0 & 0 & 0 & 0 & 0 \\ \hline
      i & 1+i & 0 & 0 & 0 & 0 & 0 & 0 & 0 & 0 & 0 & 0 & 0 & 0 & 0 & 0 \\
      -1+i & -i & 0 & 0 & 0 & 0 & 0 & 0 & 0 & 0 & 0 & 0 & 0 & 0 & 0 & 0 \\
      i & -1-i & 0 & 0 & 0 & 0 & 0 & 0 & 0 & 0 & 0 & 0 & 0 & 0 & 0 & 0 \\
      1-i & -i & 0 & 0 & 0 & 0 & 0 & 0 & 0 & 0 & 0 & 0 & 0 & 0 & 0 & 0 \\
      -i & -1+i & 0 & 0 & 0 & 0 & 0 & 0 & 0 & 0 & 0 & 0 & 0 & 0 & 0 & 0 \\
      1+i & i & 0 & 0 & 0 & 0 & 0 & 0 & 0 & 0 & 0 & 0 & 0 & 0 & 0 & 0 \\
      -i & 1-i & 0 & 0 & 0 & 0 & 0 & 0 & 0 & 0 & 0 & 0 & 0 & 0 & 0 & 0 \\
      -1-i & i & 0 & 0 & 0 & 0 & 0 & 0 & 0 & 0 & 0 & 0 & 0 & 0 & 0 & 0 \\
    \end{array}
  \right]c^{},
  \end{align}
\begin{align}
  G^{p\rm H}_0
  &= c^{\dagger}\frac{1}{4\sqrt{3}}
  \left[
    \begin{array}{cc|cccccc|cccccccc}
      0 & 0 & 0 & 0 & 0 & 0 & 0 & 0 & 0 & 0 & 0 & 0 & 0 & 0 & 0 & 0 \\
      0 & 0 & 0 & 0 & 0 & 0 & 0 & 0 & 0 & 0 & 0 & 0 & 0 & 0 & 0 & 0 \\ \hline
      0 & 0 & 0 & 0 & 0 & 0 & 0 & 0 & 0 & -i & 0 & -i & 0 & -i & 0 & -i \\
      0 & 0 & 0 & 0 & 0 & 0 & 0 & 0 & -i & 0 & -i & 0 & -i & 0 & -i & 0 \\
      0 & 0 & 0 & 0 & 0 & 0 & 0 & 0 & 0 & -1 & 0 & -1 & 0 & -1 & 0 & -1 \\
      0 & 0 & 0 & 0 & 0 & 0 & 0 & 0 & 1 & 0 & 1 & 0 & 1 & 0 & 1 & 0 \\
      0 & 0 & 0 & 0 & 0 & 0 & 0 & 0 & -i & 0 & -i & 0 & -i & 0 & -i & 0 \\
      0 & 0 & 0 & 0 & 0 & 0 & 0 & 0 & 0 & i & 0 & i & 0 & i & 0 & i \\ \hline
      0 & 0 & 0 & i & 0 & 1 & i & 0 & 0 & 0 & 0 & 0 & 0 & 0 & 0 & 0 \\
      0 & 0 & i & 0 & -1 & 0 & 0 & -i & 0 & 0 & 0 & 0 & 0 & 0 & 0 & 0 \\
      0 & 0 & 0 & i & 0 & 1 & i & 0 & 0 & 0 & 0 & 0 & 0 & 0 & 0 & 0 \\
      0 & 0 & i & 0 & -1 & 0 & 0 & -i & 0 & 0 & 0 & 0 & 0 & 0 & 0 & 0 \\
      0 & 0 & 0 & i & 0 & 1 & i & 0 & 0 & 0 & 0 & 0 & 0 & 0 & 0 & 0 \\
      0 & 0 & i & 0 & -1 & 0 & 0 & -i & 0 & 0 & 0 & 0 & 0 & 0 & 0 & 0 \\
      0 & 0 & 0 & i & 0 & 1 & i & 0 & 0 & 0 & 0 & 0 & 0 & 0 & 0 & 0 \\
      0 & 0 & i & 0 & -1 & 0 & 0 & -i & 0 & 0 & 0 & 0 & 0 & 0 & 0 & 0 \\
    \end{array}
  \right]c^{},
\end{align}
\begin{align}
  G^{\rm HH}_0 &= c^{\dagger}
  \frac{1}{4\sqrt{3}}
  \left[
    \begin{array}{cc|cccccc|cccccccc}
      0 & 0 & 0 & 0 & 0 & 0 & 0 & 0 & 0 & 0 & 0 & 0 & 0 & 0 & 0 & 0 \\
      0 & 0 & 0 & 0 & 0 & 0 & 0 & 0 & 0 & 0 & 0 & 0 & 0 & 0 & 0 & 0 \\ \hline
      0 & 0 & 0 & 0 & 0 & 0 & 0 & 0 & 0 & 0 & 0 & 0 & 0 & 0 & 0 & 0 \\
      0 & 0 & 0 & 0 & 0 & 0 & 0 & 0 & 0 & 0 & 0 & 0 & 0 & 0 & 0 & 0 \\
      0 & 0 & 0 & 0 & 0 & 0 & 0 & 0 & 0 & 0 & 0 & 0 & 0 & 0 & 0 & 0 \\
      0 & 0 & 0 & 0 & 0 & 0 & 0 & 0 & 0 & 0 & 0 & 0 & 0 & 0 & 0 & 0 \\
      0 & 0 & 0 & 0 & 0 & 0 & 0 & 0 & 0 & 0 & 0 & 0 & 0 & 0 & 0 & 0 \\
      0 & 0 & 0 & 0 & 0 & 0 & 0 & 0 & 0 & 0 & 0 & 0 & 0 & 0 & 0 & 0 \\ \hline
      0 & 0 & 0 & 0 & 0 & 0 & 0 & 0 & 0 & 0 & 0 & -1-i & -i & -1 & -i & -i \\
      0 & 0 & 0 & 0 & 0 & 0 & 0 & 0 & 0 & 0 & 1-i & 0 & 1 & i & -i & i \\
      0 & 0 & 0 & 0 & 0 & 0 & 0 & 0 & 0 & 1+i & 0 & 0 & -i & i & -i & 1 \\
      0 & 0 & 0 & 0 & 0 & 0 & 0 & 0 & -1+i & 0 & 0 & 0 & i & i & -1 & i \\
      0 & 0 & 0 & 0 & 0 & 0 & 0 & 0 & i & 1 & i & -i & 0 & 0 & 0 & 1-i \\
      0 & 0 & 0 & 0 & 0 & 0 & 0 & 0 & -1 & -i & -i & -i & 0 & 0 & -1-i & 0 \\
      0 & 0 & 0 & 0 & 0 & 0 & 0 & 0 & i & i & i & -1 & 0 & -1+i & 0 & 0 \\
      0 & 0 & 0 & 0 & 0 & 0 & 0 & 0 & i & -i & 1 & -i & 1+i & 0 & 0 & 0 \\
    \end{array}
  \right]c^{}.
\end{align}
\end{widetext}


\bibliography{main}
\end{document}